# Autocatalytic MBE growth of GaAs nanowires on oxidized Si(100)

*Janusz Sadowski[1,2*], Piotr Dłużewski[2], Janusz Kanski[3]*

(1) MAX-Lab, Lund University, P.O. Box 118, 221 00 Lund, Sweden

(2) Institute of Physics, Polish Academy of Sciences, al. Lotników 32/46, 02-668 Warszawa, Poland

(3) Department of Applied Physics, Chalmers University of Technology, SE-412 96 Göteborg, Sweden

ABSTRACT

GaAs nanowires were grown by molecular beam epitaxy on Si(100) substrates covered with 5 nm $SiO_2$. The growth was performed with $As_4$ at low (close to stoichiometric) $As_4$/Ga flux ratio, using Ga nanodroplets as catalyst. In order to produce appropriately small Ga droplets a two-stage deposition method was employed, the first leading to formation of GaAs nanocrystals on the $SiO_2$ surface, and the second leading to formation of Ga droplets at fixed positions (between and at the {111} facets of GaAs nanocrystallites). The nanowires are found to be in epitaxial relation to the Si substrate due to contact via pinholes through the $SiO_2$ layer. The NWs grow 70 times faster than the planar growth rate at the applied Ga flux. After one hour growth time the nanowires are typically up to 15 micrometers long and have diameters of about 100 nm, the catalyzing Ga nanodroplets are observed at their tops.

[*] corresponding author  e-mail: janusz.sadowski@maxlab.lu.se

The use of silicon substrates for the growth of GaAs nanowires (NW) is attractive in view of potential applications and ease of integration with existing microelectronics technology. The growth of NWs from III-V semiconductors on silicon substrates usually employs Au as catalyst, both in MBE and MOVPE growth techniques.[1,2] Here we report a self catalyzing method for NWs formation, where one of the NW components (Ga) acts as catalyst. Use of Ga instead of Au has the advantage of avoiding unintentional Au-doping, which is an undesired consequence of Au-stimulated growth procedures.[3] A limited number of studies of self-catalyzing GaAs NW growth on Si have been reported.[4,5] It has also been observed that a thin $SiO_2$ film on the Si substrate plays an important role in the NW growth process.[6] The same applies to the GaAs NWs growth on $SiO_2$-covered GaAs substrates,[7] in which case the presence of $SiO_2$ layer promotes the formation of Ga nano-droplets which can act as catalysts for GaAs NWs. On clean GaAs usually huge droplets (micrometer size) are obtained, too big to act as NW growth catalysts. In the case of $SiO_2$ coated GaAs micro-craters are formed beneath the Ga droplets, through which the NWs make contact with crystalline GaAs and become crystallographically coherent with the substrate.[8]

In our experiments the GaAs NWs were grown on Si(100) terminated with a 5 nm thick surface oxide layer. The As flux during the NWs growth was obtained from a valved cracker source (DCA), operated with the cracker stage at a low temperature to deliver predominantly $As_4$ molecules. Before the growth the $SiO_2$/Si(100) substrate was heated at 650 °C for 15-30 min at a background pressure of about $5 \times 10^{-9}$ mbar with As valve closed. During the Si substrate heating the RHEED image remained diffuse, revealing the presence of an amorphous $SiO_2$ layer.

At the end of this first heating procedure the substrate surface was covered with Ga nanodots by opening the Ga shutter for a period equivalent to that required for growth of 3 ML GaAs. Then the substrate temperature was decreased quickly to 580 °C and GaAs was grown for 10-15 min. with an $As_4$/Ga ratio of about 2. Shortly after growth start a ring pattern appeared in RHEED, indicating formation of a polycrystalline GaAs layer. No signs of NW- diffraction could be observed during this growth stage. After 10-15 min (depending on the sample) of polycrystalline GaAs deposition, both shutters and the As valve were closed, and the substrate temperature was raised to 650 °C. Upon reaching this temperature the Ga source was opened again for the same time as in the first Ga deposition stage. The sample temperature was kept at this level for another 5 minutes, and was then lowered to 580 °C for further GaAs growth with an $A_4$/Ga flux ratio of about 2. In this second stage of GaAs growth sharp streaks appeared in RHEED, slightly elongated in the direction parallel to the shadow edge of the RHEED image. These RHEED patterns are typical for nanowires.[9, 10] We believe that coalescence of nanodroplets was inhibited in this case due to the surface roughness induced by GaAs nano-crystallites.

Fig. 1 shows SEM and cross sectional TEM pictures of GaAs NWs grown on $SiO_2$/Si(100) with the two-stage Ga droplet deposition method described above. The growth was continued for 1h with a Ga flux corresponding to a planar growth rate of 0.2 ML/s. The cross-sectional TEM specimen was cut out of the sample in the (110) plane, so the NWs lie partly in the image plane, and partly in the perpendicular plane. The length of the NWs is about 15 μm, which means that the NWs growth is about 70 times faster than the planar growth. This proves that Ga nanodroplets are robust growth catalysts.

Both the SEM pictures and the RHEED observed during growth show that there is an epitaxial relation of the GaAs NWs with respect to the Si substrate: the NWs grow along the four <111> directions coming out of the Si(100) surface. Detailed analysis of the TEM pictures at the $SiO_2$/Si – NW interface region shows that NWs are contacted with Si substrate via small pin-holes. This is shown in Fig. 2a on a low-resolution and in Fig 2b on a high-resolution image. The small pinhole through the $SiO_2$ layer consists of GaAs coherently strained to the Si lattice. As seen in the low-resolution image, the GaAs microcrystallite is developed from the GaAs nanocrystal in the pinhole. This nanocrystal may contain a twin defect, as shown in Fig. 2b. The GaAs NWs with diameters of about 1000 Å then emerge from the {111} sides of the GaAs microcrystallite after the second Ga deposition.

Fig. 3 shows a TEM picture of NWs scratched off the substrate and placed on a TEM grid (a holey carbon grid). In the process of removal from SiO2/Si substrate the NWs were separated together with GaAs microcrystallite roots, which are clearly visible at the base of each nanowire. This means that the whole NWs were collected on the TEM grids, which makes it possible to determine precisely their lengths. A TEM image of a top part of an individual NW is shown in Fig.4. At its end the Ga catalyst is clearly visible in the form of nano-ball. The NW has a high density of stacking faults, which is typical for GaAs NWs grown along the <111> direction.

In summary, we have demonstrated self-catalyzed epitaxial growth of GaAs nanowires on oxidized Si(100) substrate using Ga nanodroplets as catalysts. The Ga droplets were deposited in two stages, the first leading to development of high density GaAs microcrystallites on the $SiO_2$ surface, the second catalyzing the NWs growth. We have shown that GaAs NWs are in epitaxial relation to the Si(100) substrate due to GaAs nanocrystals penetrating the SiO2 layer. To our knowledge this is the first demonstration of Ga-catalyzed MBE growth of epitaxial GaAs nanowires on oxidized silicon substrates.

ACKNOWLEDGEMENTS. This work was partially supported by the Ministry of Science and Higher Education (Poland) through grant N N202 126035 and by the Swedish Research Council (VR). The authors would like to thank A. Presz from Institute of High Pressure Physics (UNIPRESS), Polish Academy of Sciences, Warsaw, for SEM measurements.

FIGURE CAPTIONS

**Fig.1.** (a) - SEM and (b) - (011) plane cross sectional TEM images of GaAs NWs grown on Si(100) substrate with 5 nm surface $SiO_2$ layer.

**Fig.2.** Low resolution (a), and high resolution (b) cross sectional TEM images from GaAs microcrystal at the base of GaAs NW. GaAs microcrystal is separated from the Si(100) substrate by 5 nm thick $SiO_2$ oxide layer. The microcrystal is contacted with Si(100) substrate by a small (20 nm in diameter) nano-pinhole filled with monocrystalline GaAs coherently strained to silicon. The twin defect is visualized in the GaAs microcrystal grown around a pinhole nanocrystal.

**Fig.3.** TEM images from GaAs NWs scratched out off the Si(100) substrate and placed on a holey carbon TEM grid.

**Fig.4.** TEM image of the top part of GaAs NW with catalyzing Ga nano-droplet at the end. Stacking faults are clearly visible, starting at the distance of about 20 nm from the GaAs-Ga interface.

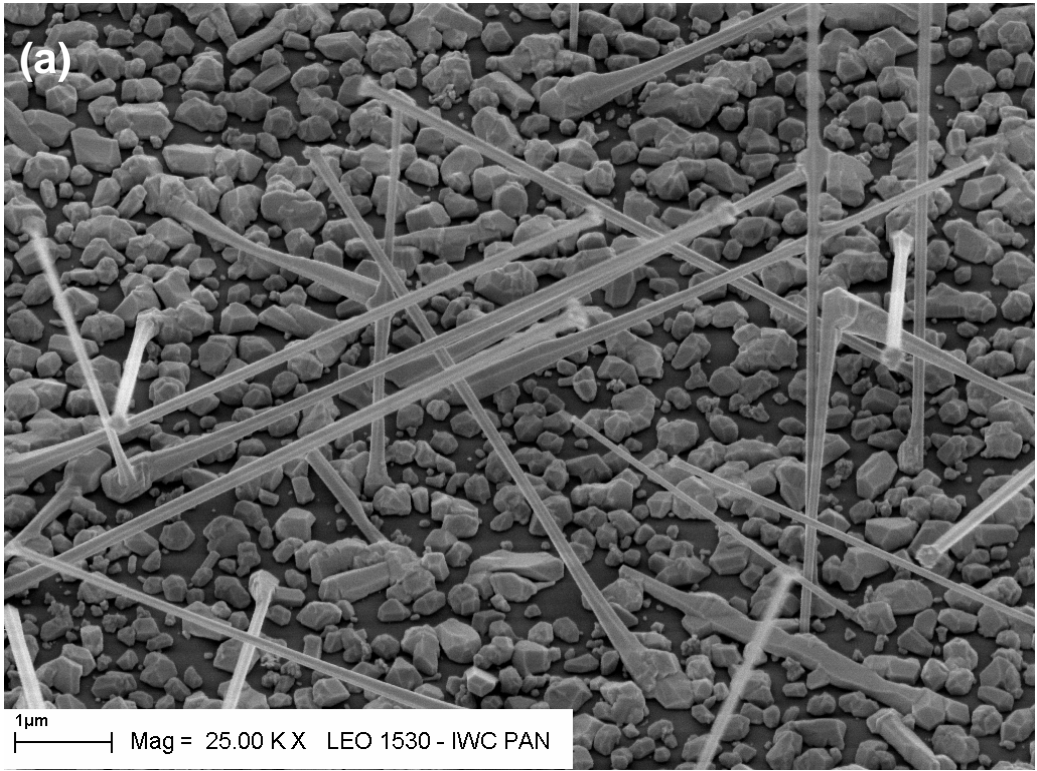

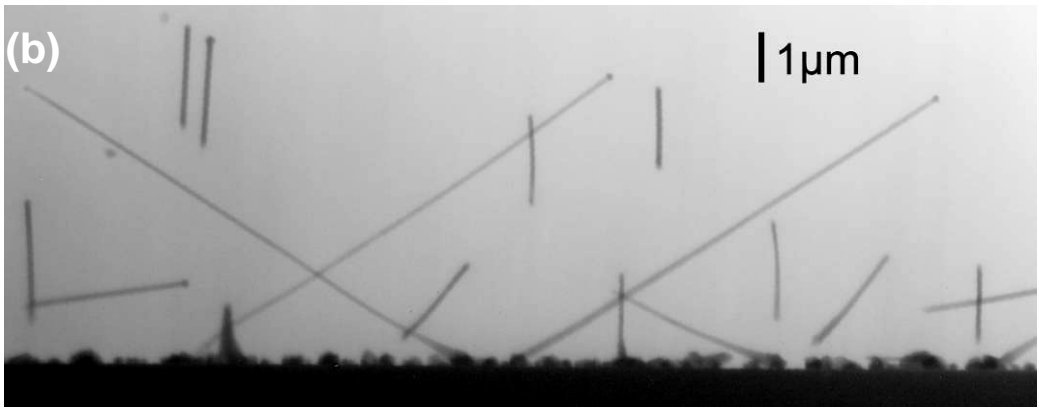

Fig.1

Fig.2

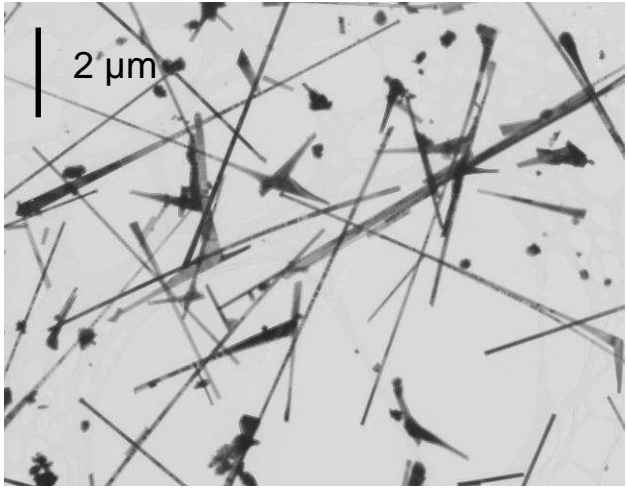

Fig.3

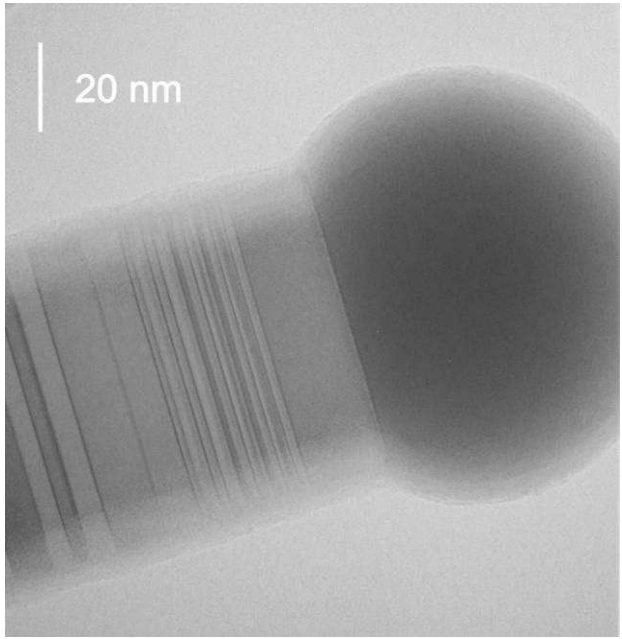

Fig.4